\documentclass[onecolumn,floatfix,showpacs,nofootinbib,11pt]{revtex4-1}
\usepackage{graphicx,upgreek}
\usepackage{amssymb,amsmath}
\usepackage{graphicx,float,color}
\usepackage{indentfirst,makeidx}
\def\is{\!=\!}
\newcommand{\x}{X}
\usepackage[colorlinks,hyperindex]{hyperref}

\definecolor{green1}{RGB}{0,128,0} 
\hypersetup{hidelinks,backref=true,pagebackref=true,hyperindex=true,colorlinks=true,breaklinks=true,urlcolor= blue}
\hypersetup{%
  colorlinks = true,
  linkcolor  = blue,
  citecolor = green1,
}
\usepackage{bookmark,textgreek}
\usepackage{hyperref,color,xcolor}
\hypersetup{hidelinks,hyperindex=true,colorlinks=true,breaklinks=true,urlcolor= blue}
\hypersetup{%
  colorlinks = true,
  linkcolor  = blue
}

\def\da{\delta_A}
\def\db{\delta_B}
\newcommand{\bege}{\begin{equation}}
\newcommand{\enge}{\end{equation}}
\newcommand{\vv}{{\bf v}}

\newcommand{\cv}{\circ}
\usepackage{mathtools,harmony}
\newcommand{\ZZ}{\mathbb{Z}}

\newcommand{\noi}{\noindent}
\newcommand{\ee}{{e}}
\usepackage{color,accents,stmaryrd}
\newcommand{\op}{\oplus}
\usepackage{mathrsfs}
\usepackage{amssymb}
\usepackage{amsmath}
\usepackage{dsfont}
\usepackage{indentfirst}
\newcommand{\beq}{\begin{eqnarray}}\newcommand{\benu}{\begin{enumerate}}\newcommand{\enu}{\end{enumerate}}
\newcommand{\eeq}{\end{eqnarray}}
\newcommand{\oct}{\mathbb{O}}
\newcommand{\mani}{M}

\usepackage{latexsym}
\newcommand{\cl}{{\mathcal{C}}\ell}
\newcommand{\la}{\Lambda}
\begin{document}
\title{Additional fermionic fields onto parallelizable 7-spheres}
\author{A. Yanes}
\email{aquerman.m@ufabc.edu.br}
\affiliation{Centro de Matem\'atica, Computa\c c\~ao e Cogni\c c\~ao, Universidade Federal do ABC, 09210-580,
Santo Andr\'e, Brazil.}

\author{R. da~Rocha}
\email{roldao.rocha@ufabc.edu.br}
\affiliation{Centro de Matem\'atica, Computa\c c\~ao e Cogni\c c\~ao, Universidade Federal do ABC, 09210-580,
Santo Andr\'e, Brazil.}
\begin{abstract}
The geometric Fierz identities are here employed to generate new emergent fermionic fields on the parallelizable (curvatureless, torsionfull) 7-sphere ($S^7$). Employing recently found new classes of spinor fields on the $S^7$ spin bundle, new classes of fermionic fields are obtained from their bilinear  covariants by a generalized reconstruction theorem, on the parallelizable $S^7$.  Using a generalized non-associative product on the octonionic bundle on the parallelizable $S^7$, these new classes of algebraic spinor  fields,  lifted onto the parallelizable  $S^7$, are shown to 
correctly transform under the Moufang loop generators on $S^7$.
  \end{abstract}
\maketitle
\section{Introduction}

(Classical) spinor fields are well known to be elements in the carrier space 
of the Spin group irreducible representations on any given spacetime that admits a spin structure, namely, if the second Stiefel--Whitney class vanishes. Spinor fields are, in particular,  employed for constructing the so called bilinear covariants, consisting of tensorial quadratic forms involving the spinors. The bilinear covariants were shown to be 
the homogeneous part of a multivector Fierz aggregate \cite{BBR}.
Particularizing for 4D the Minkowski spacetime case,  spinor fields were  
 classified with respect to their bilinear covariants in the Dirac--Clifford algebra, by the so called Lounesto's spinor field classification \cite{lou2}. Further lattice generalizations in the context of quantum Clifford algebras were also studied \cite{Ablamowicz:2014rpa}. The bilinear covariants are not independent, but constrained by 
the Fierz identities \cite{lou2,Fabbri:2016msm}. Reciprocally, 
given the bilinear covariants, their associated spinor fields can be re-obtained up to a phase, by the reconstruction theorem  \cite{Mosna:2003am,Cra}. The Lounesto's classification is based upon the U(1) gauge  
symmetry of the first-order equations of motion that rule
spinor fields in each spinor class. However, a more general classification has been proposed in Ref. \cite{Fabbri:2017lvu} encompassing spinor multiplets as realizations of (non-Abelian) gauge fields. In this 
more general classification, composed flagpoles,
dipoles, and flag-dipoles naturally descend within fourteen disjoint classes of spinor fields, under the gauge symmetry SU(2) $\times$ U(1). In this setup, the spinor fields in the standard Lounesto's classification were shown to be a limiting case,  equivalent to Pauli singlets \cite{Fabbri:2017lvu}. 
Further spinor representations were studied in Refs. \cite{123,HoffdaSilva:2017vic}, with also 
other proposals to construct the bilinear covariants for flagpole spinors \cite{HoffdaSilva:2016ffx}. 
An analogous
classification in the framework of second quantization and a quantum reconstruction algorithm was also proposed, being the Feynman propagator 
extended for regular and singular spinor fields, in Ref. \cite{Bonora:2017oyb}.

In any fixed spacetime dimension, $n$, and signature, $(p,q)$, the very construction of 
the bilinear covariants depends on the existence of either real, or complex, or even quaternionic structures. Hence,  the existence of non null bilinear covariants can  be impeded by the geometric Fierz identities. 
Despite the natural obstructions due to the existence of algebraic and geometric structures on a given spacetime dimension/signature, the Lounesto's spinor field classification on 4D Minkowski spacetime was successfully generalized to other spacetime 
dimensions and signatures, of relevance in their applications, as the  emergence of fermionic fields in the respective spacetime compactifications. 
Spinor fields on the 7-sphere $S^7$, as an Einstein space composing the compactification AdS$_4\times S^7$, were studied in Ref. \cite{BBR}, where new spinor classes were derived. On the other hand, new spinor field classes in the  compactification AdS$_5\times S^5$ were derived and investigated in Ref. \cite{brito}, representing new recently obtained fermionic solutions in string theory.  
More precisely, Ref. \cite{BBR} proposed new classes of spinor fields on $S^7$, based on the geometric Fierz identities in Ref. \cite{1}. The underlying structure of the geometric Fierz identities on $S^7$ was shown to sternly obstruct the amount of non null bilinear covariants were found on $S^7$. Nevertheless, further three new emergent classes of fermionic fields on $S^7$. From a more physical point of view, investigating 
these new classes of spinors $S^7$ may afford new fermionic solutions of first order equations of motion, that can play an important role on 
supergravity. In fact, 
one of the spontaneous compactification schemes on $n = 11$ supergravity can be implemented by the so called Freund--Rubin--Englert 
solution, obtained on a product manifold AdS$_4\times S^7$ \cite{Gursey:1983yq}. As important as the standard $S^7$, the so called parallelizable $S^7$, a curvatureless manifold that has torsion, emerges when the antisymmetric gauge field strength in the Englert's solution excedes the Freund--Rubin one, 
being identified with the Cartan--Schouten torsion on the 7-sphere.

Our main aim here is to construct new fermionic fields on the parallelizable $S^7$, that can be then obtained when new classes of $S^7$ spinor fields are lifted onto the parallelizable  $S^7$. 
This paper is organised as follows: in Sect. II, after briefly reviewing how the geometric Fierz identities 
are used to derive additional spinor field classes on $S^7$, we propose a reconstruction procedure for obtaining the spinor fields, in these new classes, from the bilinear covariants and the geometric Fierz identities. Sect. III is then devoted to briefly review the parallelizable sphere, whose torsion is defined with respect to the non-associative $X$-product on the octonionic bundle. The geometric Fierz identities 
are used to derive the spinor field classes on $S^7$, that are going to be lifted onto the parallelizable $S^7$, whereon new fermionic fields can be then constructed through the introduction of a generalized octonionic law of transformation.

\section{Geometric Fierz Identities and Bilinear Covariants}
Let $(M,g)$ be a manifold endowed with a metric tensor. The exterior bundle $\Upomega(M)=\oplus_{i=0}^\infty\Upomega_i(M)$ 
has endomorphisms that come from the tensor algebra quotient construction. Given a $k$-form field\footnote{{\color{black}{One calls a $k$-form field a section of an homogeneous space of the exterior bundle.}}} $a \in \sec \Upomega^k(M)$, the grade 
involution, $\hat{a}=(-1)^{k}a$, is an automorphism; the reversion, $\tilde{a}=(-1)^{[(k/2)]}a,$ for $[(k)]$ denoting the integer part of the degree $k$, is an antiautomorphism. These composition of these two morphisms define the conjugation, denoted  by $\bar{a}$. The Clifford bundle 
can be obtained by equipping the exterior bundle  with the universal Clifford product  
$u \diamond  a = u \wedge a+ u 
\lrcorner a$, for all 1-forms $u \in \sec\Upomega^1(M)$,  where $\lrcorner$ is the left contraction. 

The spinor bundle of the Minkowski spacetime $\mathbb{R}^{1,3}$ is composed by spinor fields, $\psi$, carrying the
${\left(\frac12,0\right)}\oplus{\left(0,\frac12\right)}$ representations of the Lorentz group.  The bilinear covariants are sections of the exterior bundle $\Upomega(M)$. With respect to a {basis $\{e^\mu\}$}, they read \begin{subequations}
\begin{eqnarray}
\upsigma &=& \bar{\psi}\psi\in\sec\Upomega^0(M)\,,\label{sigma}\\
\textsc{J}&=&\textsc{J}_{\mu}e^{\mu }\in\sec\Upomega^1(M)\,,\label{J}\\
\textsc{S}&=&S_{\mu\nu }e^{\mu}\wedge e^{ \nu }\in\sec\Upomega^2(M)\,,\label{S}\\
\textsc{K}&=& K_{\mu }e^{\nu }\in\sec\Upomega^3(M)\,,\label{K}\\\omega&=&\bar{\psi}\gamma_{0}\gamma_1\gamma_2\gamma_3\psi\in\sec\Upomega^4(M)\,,  \label{fierz.}
\end{eqnarray}\end{subequations}
where $\textsc{J}_{\mu}=\bar{\psi}\gamma _{\mu }\psi$, $S_{\mu\nu }=\bar{\psi}\upsigma _{\mu
\nu }\psi$, $K_{\mu }=i\bar{\psi}\gamma_{0}\gamma_1\gamma_2\gamma_3\gamma _{\mu }\psi
\,,$ are the respective components in Eqs. (\ref{J}) -- (\ref{K});  
$\gamma_5:=i\gamma_0\gamma_1\gamma_2\gamma_3$ and $\bar\psi=\psi^\dagger\gamma_0$. Besides, $\upsigma_{\mu\nu}:=\frac{i}{2}[\gamma_\mu, \gamma_\nu]$. Gamma matrices satisfy a Clifford algebra {\color{black}{named}} $\cl_{1,3}$, $\gamma_{\mu }\gamma _{\nu
}+\gamma _{\nu }\gamma_{\mu }=2g_{\mu \nu }\mathbf{1}$, {where $g_{\mu\nu}$ denotes the 
Minkowski spacetime metric components.}
When not both $\upsigma$ and $\omega$ vanish altogether, the bilinear covariants are governed by the Fierz identities \cite{lou2} 
\begin{equation}\label{fifi}
\textsc{K}^2+{\rm J}^2
=0={\rm J}\cdot\textsc{K},\qquad(\omega+\upsigma\gamma_{0}\gamma_1\gamma_2\gamma_3)\textsc{S}={\rm K}\wedge\textsc{J},\qquad
\omega^{2}+\upsigma^{2}={\rm J}^2\,.  
\end{equation}
\noindent  

Lounesto derived, from the bilinear covariants, a classification of spinor fields \cite{lou2}, for  
${\rm J}\neq0$. However, this condition that was firstly motivated by the Dirac electron theory, and can be circumvented in three additional classes that were recently derived in Minkowski spacetime \cite{EPJC}, conjectured to consist of ghost spinors. Apart from these classes of ghost spinors, the original Lounesto's classification splits the spinor fields on Minkowski spacetime into 
six disjoint classes. In Eqs. (\ref{ppo}) -- (\ref{ppl}) below, we just denote the bilinear covariants that do not vanish:
\begin{subequations}
\begin{eqnarray}
&&1)\;\;\omega\neq0,\;\;\;  \upsigma\neq0,\;\;\;\textsc{K}\neq 0, \;\;\;\textsc{S}\neq0\;\; \\  
\label{ppo}
&&2) 
\;\;\upsigma\neq0,\label{dirac1}\;\;\;\textsc{K}\neq 0, \;\;\;\textsc{S}\neq0\;\;\;\\   
&&3)\;\;\omega \neq0, \;\;\;\textsc{K}\neq 0\;\;\;\textsc{S}\neq0 \;\;\;\;\\  
&&4)\;\;\textsc{K}\neq 0, \;\;\;\textsc{S}\neq0  \;\;\quad\qquad\\
&&5)\;\;\textsc{K}=0,\;\;\; \textsc{S}\neq0 \;\; \quad\qquad\\
&&6)\;\;\textsc{K} \neq 0,\;\;\; \textsc{S}=0
\;\;\;\quad\qquad\label{ppl}
\end{eqnarray}
\end{subequations} Classes 1, 2, and 3 consist of regular spinor fields, since not both the scalar and the pseudoscalar vanish. Classes 4, 5, and 6 realize singular spinor fields, where both $\upsigma$ and $\omega$ are null. Refs. \cite{Ablamowicz:2014rpa,esk,EPJC,Cavalcanti:2014uta} illustrate a vast range of applications of these classes in quantum field theory and gravity. 
Ref. \cite{Fabbri:2016msm} introduced more two exclusive classes into the Lounesto's classification, through a generally-relativistic gauge classification, whereas the most general spinor field class in each spinor class was derived 
in Ref. \cite{Cavalcanti:2014wia} as a prominent computational tool for the reconstruction theorem.

 The Fierz identities (\ref{fifi}) are well known not to be valid for the case of singular spinors. In this case, based upon a Fierz aggregate, 
 \begin{eqnarray}
{\rm Z}= \frac12(\omega\gamma_{0}\gamma_1\gamma_2\gamma_3+i\textsc{K}\gamma_{0}\gamma_1\gamma_2\gamma_3+i\textsc{S}+ {\rm J} +\upsigma) \,, \label{Z}\label{zsigma}
\end{eqnarray}
 the Fierz identities (\ref{fifi}) can be replaced  by 
 \begin{subequations}\beq
\label{nilp}{\rm Z}^{2}{}  &=&\upsigma{} {\rm Z}{},\\
{\rm Z}{}\gamma_{\mu}{\rm Z}{}&=&J_{\mu}{}{\rm Z}{},\\
{\rm Z}{}\sigma_{\mu\nu}{\rm Z}{}&=&S_{\mu\nu}{}{\rm Z}{},\\
{\rm Z}{}i\gamma_{0}\gamma_{1}\gamma_{2}\gamma_{3}\gamma_{\mu}{\rm Z}{}  &=&K_{\mu}{}{\rm Z}{},\\
- {\rm Z}{}\gamma_{0}\gamma_{1}\gamma_{2}\gamma_{3}{\rm Z}{}&=&\omega{} {\rm Z}{}.\label{nilp1}
\eeq\end{subequations}
Fierz aggregates that are self-adjoint multivectors, $\gamma_{0}{\rm Z}\gamma_{0}={\rm Z}^{\dagger}$, are better known as boomerangs  \cite{lou2}.  

Given any spinor $\upupsilon\in\mathbb{C}^4$ such that $\bar\upupsilon\gamma_0\psi\neq 0$, the non trivial spinor $\psi$ can be, then,  reconstructed by the inversion theorem, as $
\psi=\frac{1}{2\sqrt{\bar\upupsilon\mathbf{Z}\upupsilon}}\;e^{-i\alpha}\mathbf{Z}\upupsilon, \label{31}%
$  for an arbitrary phase $\alpha$, such that 
$-i\alpha=\ln\left({2}\sqrt{\bar\upupsilon\psi\bar\upupsilon\mathbf{Z}\upupsilon}\right)$. In particular, any regular spinor can be   reconstructed as \cite{Cra,Mosna:2003am}
\beq
\psi = \frac12\sqrt{J_0+\sigma - K_3 + S_{12}}\; {\rm Z} e^{i\alpha}(1,0,0,0)^\intercal.
 \eeq
 
{\color{black}{ Heretofore spinor fields were approached without mentioning 
 the spinor bundle. We denoted the Minkowski spacetime manifold by $M\simeq \mathbb{R}^{1,3}$.
 Since it is an affine space, being isomorphic to its own tangent spaces, a lot of important structures were hidden throughout the text, for simplicity. Nevertheless, to approach spinor fields on higher dimensions, we should recall the spinor structures of Minkowski spacetime.  Spinor fields are  sections of the so called spinor bundle. For defining it, some underlying  structures are introduced in the Appendix \ref{app}.
 }}

{\color{black}{Now, to}} define and construct analogous classifications on spacetimes of any dimension and signature, when it is possible, the geometric Fierz identities can be analyzed when a spin structure endows an $M$ manifold. For it, the so called K\"ahler-Atiyah bundle introduced, which consists of the exterior bundle endowed with the Clifford product, denoted in this section by $\diamond$.{\color{black}{ Considering our case of interest, consisting of the 7-sphere $S^7$}}, its $S$ spin bundle is, thus, equipped with the induced $\Ganz: S\to S$ product, accordingly  
\cite{1}. {\color{black}{This composition just indicates the product between spinor fields, usually denoted by juxtaposition, when Minkowski spinor fields are regarded.}} {\color{black}{Denoting by ${\rm End}(S)$ all 
the linear mappings from $S$ to $S$ and by ``sec'' any section of a bundle}}, a  bilinear pairing $B: \sec S\times \sec S\to\mathbb{R}$ can define a bilinear mapping \cite{1,BBR}. 
Indeed,  given sections $\psi,\uppsi$ on the spin bundle, a  bilinear mapping  $B_0: \sec S\times \sec S\to \mathbb{R}$, on the $S^7$ spinor bundle, reads \cite{1} 
\begin{eqnarray}
\!\!\!\!\!\!B_0(\psi,\uppsi)\!=\! B\!\left(\!{}_{\Re}\psi,\!{}_{\Re}\uppsi\right)\!-\!
B\!\left(\!{}_{\Im}\psi,\!{}_{\Im}\uppsi\right)\!+\!i\!\left[B\!\left(\!{}_{\Re}\psi,\!{}_{\Im}\uppsi\right)\!+\! B\!\left(\!{}_{\Im}\psi,\!{}_{\Re}\uppsi\right)\right],\label{formaa}
\end{eqnarray}\noindent  {\color{black}{for the real, ${}_{\Re}\psi$,  and the imaginary, ${}_{\Im}\psi$}},  components of the spinor field $\psi$ \cite{1}. 
{\color{black}{This bilinear mapping is the one that shall generalize the bilinear covariants (\ref{ppo} - \ref{ppl}) scalar components, that were constructed on $\mathbb{R}^{1,3}$ to the 7-sphere. This can be 
implemented by the bilinear mapping on the {\color{black}{$S^7$}} spin bundle: }}
\begin{eqnarray}\label{formab}
B_k(\psi,\uppsi)=B(\psi,\gamma_{\tau_1}\dots\gamma_{\tau_k} \uppsi)=
{\bar{\psi}}{\gamma}_{\tau_1}\dots\gamma_{\tau_k} {\uppsi}\,.
\end{eqnarray}\noindent  

To define new spinor classes on $S^7$, when $k$ is odd, the bilinear mapping $
  B (\psi,\gamma^{\tau_1}\ldots\gamma^{\tau_k}\uppsi)$ is not equal to zero \cite{1}.
Defining 
\begin{equation}
\mathcal{A}_{\uppsi |\Uppsi}(\psi):= B (\psi,\Uppsi)\uppsi\,,\quad \text{for all}\;\;\;\;\;
\psi,\uppsi,\Uppsi \in \sec S\,,
\end{equation}\noindent  given $\mathring\psi,\mathring\uppsi,\mathring\Uppsi \in \sec S$, then 
the (geometric) Fierz identities then read \cite{1}
\begin{equation}
\mathcal{A}_{\uppsi |\Uppsi}\,\Ganz\, \mathcal{A}_{\mathring\uppsi |\mathring\Uppsi}= B
(\mathring\uppsi,\Uppsi)\mathcal{A}_{\uppsi |\mathring\Uppsi}\,.
\end{equation} 
{\color{black}{Given the structure $D$ that defines the complex conjugate on $S$ by $D(\psi)={}_{\Im}\psi$}}, the elements $\mathcal{A}_{\psi |\uppsi}$ {\color{black}{are differential forms that}} can be always split into  $
\mathcal{A}_{\psi |\uppsi}=D\,\Ganz\, \mathcal{A}^1_{\psi |\uppsi}+\mathcal{A}^0_{\psi |\uppsi}$ \cite{1}, 
where
\begin{subequations}
\begin{eqnarray}
 \mathcal{A}^{\psi |\uppsi} &=&\sum_{k=0}^7\frac{1}{k!}(-1)^{k}    B (\psi,
\gamma_{\tau_1}\ldots\gamma_{ \tau_k}\uppsi)e^{\tau_1}\wedge\cdots\wedge e^{\tau_k}\,,\label{eoo1} \\
 \mathcal{A}_D^{\psi |\uppsi} &=&\sum_{k=0}^7    \frac{(-1)^{k}}{k!}
 B (\psi, D\,\Ganz\, \gamma_{\tau_1}\ldots\gamma_{
\tau_k}\uppsi)e^{\tau_1}\wedge\cdots\wedge e^{ \tau_k}\,. \label{eoo2}
\end{eqnarray}
\end{subequations}

The geometric Fierz identities then {\color{black}{follow}} for $S^7$ \cite{1}:
\begin{subequations}\begin{eqnarray}
 \hat{\mathcal{A}}^{\uppsi |\Uppsi}\diamond   \mathcal{A}_D^{\mathring\uppsi |\mathring\Uppsi} +    
\mathcal{A}_D^{\uppsi |\Uppsi}\diamond   \mathcal{A}^{\mathring\uppsi |\mathring\Uppsi} 
&=&   B (\mathring\uppsi,\Uppsi) \mathcal{A}_D^{\uppsi |\mathring\Uppsi}\,,\label{fp1}\\
  \mathcal{A}^{\uppsi |\Uppsi}\diamond   \mathcal{A}^{\mathring\uppsi |\mathring\Uppsi} + (-1)^{k}
\hat{\mathcal{A}}_D^{\uppsi |\Uppsi}\diamond   \mathcal{A}_D^{\mathring\uppsi |\mathring\Uppsi} 
&=&   B (\mathring\uppsi,\Uppsi) \mathcal{A}^{\uppsi |\mathring\Uppsi}\,.\label{fp2}
\end{eqnarray}\end{subequations}
These equations are the equivalent of Eqs. (\ref{fifi}), for $S^7$. 

{\color{black}{Moreover, the bilinear covariants on $S^7$ emulate the ones of Minkowski spacetime (\ref{ppo} - \ref{ppl}), by   
 \begin{eqnarray}
\label{ddd}
\upphi_k= \frac{1}{k!}B (\psi,\gamma_{\tau_1}\ldots\gamma_{
\tau_k}\psi)e^{\tau_1}\wedge\cdots\wedge e^{\tau_k} \end{eqnarray}\noindent It is worth to emphasize that the bilinear covariants construction on $\mathbb{R}^{1,3}$ are not obstructed by a dimensional accident. However, on $S^7$ (and also on other specific dimensions), the geometric Fierz identities (\ref{fp1}, \ref{fp2}) severely obstruct the very existence of homogeneous bilinear covariants \cite{1}. In fact, }}
spinors on $S^7$ have the bilinear covariants $\upphi_k$ equal to zero, with the exceptions  when  $k=0$ or $k=4$ \cite{1,BBR}, namely, \beq
 \upphi_0&=& B  (\psi,\psi),\\
\label{phi4}
\upphi_4&=&\frac{1}{4!}  B (\psi,\gamma_{\tau_1}\gamma_{\tau_2}\gamma_{\tau_3}\gamma_{
\tau_4}\psi)\;e^{\tau_1}\wedge e^{\tau_2}\wedge e^{\tau_3}\wedge e^{\tau_4}\,.
\end{eqnarray}

Then, the geometric Fierz identities yield a single class Majorana  spinors
on $S^7$, given by $\upphi_0\neq 0$ and $\upphi_4\neq 0$, being all other bilinears $\upphi_k=0$, for $\{k\}\neq \{0,4\}$. 
 A higher order generalization of Eq. (\ref{formab}) is then necessary, to 
 encompass new classes of fermionic fields on $S^7$:
 \begin{eqnarray}\label{formac}
\upbeta_k(\psi,\gamma_{\tau_1}\ldots\gamma_{\tau_k}\uppsi)&=& B\left({}_{\Re}\psi,\gamma_{\tau_1}\ldots\gamma_{\tau_k}{}_{\Re}\uppsi)-
B\left({}_{\Im}\psi,\gamma_{\tau_1}\ldots\gamma_{\tau_k}{}_{\Im}\uppsi\right)\right)\\&&\nonumber\qquad\qquad+i\left[
B\left({}_{\Re}\psi,\gamma_{\tau_1}\ldots\gamma_{\tau_k}{}_{\Im}\uppsi\right)+B\left({}_{\Im}\psi,\gamma_{\tau_1}\ldots\gamma_{\tau_k}{}_{\Re}\uppsi\right)\right]\,.\end{eqnarray}\noindent 
Hence,  the complex bilinear covariants can be defined \cite{BBR},  
\beq
\Upphi_k=
\frac{1}{k!}\upbeta_k(\psi,\gamma_{\tau_1}\ldots\gamma_{\tau_k}\psi)e^{
\tau_1}\wedge\cdots\wedge e^{\tau_k},\eeq
 yielding three (non-trivial)  
classes  of spinor fields on the $S^7$ spin bundle \cite{BBR}, 
\begin{subequations}
\begin{eqnarray}
&1)\;\;\;\;\; \Upphi_0=0,\quad\Upphi_4\neq0,\label{c12}\\
&2)\;\;\;\;\; \Upphi_0\neq0,\quad\Upphi_4=0,\label{c13}\\
&3)\;\;\;\;\; \Upphi_0\neq0,\quad\Upphi_4\neq0\,.\label{c14}\end{eqnarray}
\end{subequations} 

{\color{black}{The Fierz aggregate (\ref{zsigma}) in the $\mathbb{R}^{1,3}$ Minkowski spacetime can be now emulated for the 7-sphere. In fact, }}
the reconstruction theorem can be then employed for constructing the original spinor field as a section of the spin bundle, from the corresponding Fierz aggregate 
\beq\label{aggg}
\check{\rm Z}=\Upphi_0+\Upphi_4,
\eeq \noindent that is simpler than its 4D Minkowski counterpart Fierz aggregate, defined in Eq. (\ref{zsigma}). Hence, when an arbitrary spinor $\xi\in S^7$ satisfies $\xi^{\dagger}(\upsigma_2\otimes\gamma_{0})\psi\neq0$, where $\sigma_2=\scriptsize{\begin{pmatrix}0&-i\\i&0\end{pmatrix}}$,  the original $S^7$ spinor $\psi$ can be obtained from its Fierz aggregate (\ref{aggg}),
\begin{equation}
\psi=\frac{1}{2}\left({{\langle\xi^{\dagger}(\upsigma_2\otimes\gamma_{0})\check{\rm Z}\xi}\rangle_0}\right)^{-1/2}\;e^{-i\theta}\check{\rm Z}\xi, \label{3}%
\end{equation}
\noindent where
$e^{-i\theta}={2}(\langle{\xi^{\dagger}(\upsigma_2\otimes\gamma_{0})\check{\rm Z}\xi}\rangle_0)^{-1/2}\langle\xi^{\dagger}(\upsigma_2\otimes\gamma_{0})\psi\rangle_0$.

 \section{Lifting new spinor fields on the parallelizable $S^7$}
 
 Heretofore, new classes of $S^7$ spinors were derived into the classes (\ref{c12} -- \ref{c14}), whose representative spinor fields can be reconstructed by  Eq. (\ref{3}). These representative spinor fields 
are now aimed to be lifted onto the so called parallelizable $S^7$, that  can be regarded as the manifold of unit octonions. Among the parallelizable spheres, $S^7$ is the sole one that 
 does not carry a Lie group structure, however a Moufang loop structure, instead. The (sub)bundle of  octonionic sections on $S^7$ presents a Moufang loop underlying structure, which is fiberwise. 
 From a geometric point of view, the $S^7$ algebra is a natural stage to generalize the concept of a Lie algebra, wherein the
 structures constants are substituted by the parallelizable torsion.  

{\color{black}{The octonionic algebra $\mathbb{O}$ 
is constituted by a 8-dimensional vector space, with basis $\{e_0,\ldots,e_7\}\subset \mathbb{R}^8$, with $e_0^2=1$, $e_a^2=-1$, for $a=1,\ldots,7$,
endowed with the octonionic multiplication, denoted by ``$\circ$'', which is ruled by $\ee_a\circ \ee_b=f^{c}_{ab}\ee_c-\delta_{ab}$, 
where $f^{c}_{ab}=1$ for the cyclic permutations
$\{(abc)\}=\{(126),(237),(341),(452),(563),(674),(715)\}.$ Every octonion $X\in\mathbb{O}$ can be, thus, written as $X=X_0e_0+\sum_{a=1}^7X^ae_a$. 
Instead of the vector space $\mathbb{R}^8$, one can take the paravector space}} $V_7:=\mathbb{R}\op\mathbb{R}^{0,7}$ endowed with the octonionic standard product  $\circ \colon V_7  \times V_7  \to V_7$. In fact, the scalar part, {\color{black}{$X^0$}} does correspond to the real part of an octonion, whereas  the vector component, {\color{black}{$\sum_{a=1}^7X^ae_a$, }} regards the imaginary part. 
{\color{black}{In this case, the}} identity $\ee_0=1$ and an orthonormal basis $\left\{\ee_a\right\}^{7}_{a=1}$ of  $V_7 \hookrightarrow \cl_{0, 7}$ generate the octonion algebra \cite{baez}. The octonionic product can be emulated at the Clifford algebra $\cl_{0, 7}$ as 
\beq \label{201}A\circ B=\left\langle AB(1-\mho)\right\rangle_{0\op 1}, \quad A,B \in V_7 ,\eeq
where $\mho=\ee_7\ee_1\ee_5+\ee_6\ee_7\ee_4+\ee_5\ee_6\ee_3+\ee_4\ee_5\ee_2+\ee_3\ee_4\ee_1+\ee_2\ee_3\ee_7+\ee_1\ee_2\ee_6$ {\color{black}{is a 3-form}}, and the juxtaposition denotes the Clifford product. The symbol $\langle \chi\rangle_{0\op 1}$ denotes 
the projection of a multivector $\chi\in\cl_{0,7}$ onto its paravector components. 
For the underlying Lie algebra $\mathfrak{g}_7$, the Lie bracket satisfies 
\beq
[[\ee_{i},\ee_{j}],\ee_k]+[[\ee_{k},\ee_{i}],\ee_j]+[[\ee_{j},\ee_{k}],\ee_i] 
 = (\delta_{i[k|}\delta_{j|p]}+\upepsilon_{mij}\upepsilon_{mkp})\ee_{p},\eeq\noi where $A_{[ab]}=\frac12(A_{ab}-A_{ba})$, for any tensor $A_{ab}$, and the Einstein's summation convention is used hereon. The (Clifford) conjugation of $X=X^0+X^b\ee_b \in \mathbb{O}$ reads $\bar{X}=X^0-X^b\ee_b$, for $X^0, X^a$  real coefficients.  Given $X \in S^7$, the \textit{$X$-product} is defined by \cite{ced} 
\bege\label{205}
A\circ_X B:=(A\circ X)\circ(\bar{X}\circ B).
\enge
The expressions below are shown in, e.g., \cite{ced}
\bege\label{3p14}
(A\circ X)\circ(\bar{X}\circ B)=X\circ ((\bar{X}\circ A)\circ B)=(A\circ(B\circ X))\circ \bar{X}.
\enge

As we dealed with bundles in the previous sections, the octonion bundle \begin{equation}\mathbb{O}S^7\simeq (\mathbb{R}\times S^7)\oplus TS^7,\end{equation} {\color{black}{shall be employed,}} where $TS^7$ denotes the tangent bundle on $S^7$, with fibers $\mathbb{R}\oplus T_X S^7$ \cite{Grigorian:2015jwa}. {\color{black}{Hence, given $A,B,C\in\mathbb{O}S^7$, and }}the associator $
[A,B,C] = A\circ (B\circ C)- (A\circ B)\circ C,$ 
one can write \cite{Grigorian:2015jwa}
\bege\label{205}
A\circ_X B=A\circ B + [A,B,\bar{X}]\circ X.
\enge
Although $A\cv_X B\neq A\cv B$ in general,  choosing $X$ as being an element of  the following sets of vector fields
$\{\pm \ee_{b}\}$, $\{(\pm\ee_{a}\pm \ee_{b})/\sqrt{2}\}$, $\{(\pm \ee_{a}\pm \ee_{b}\pm \ee_{c}\pm \ee_{d})/2\,|\,  \; \ee_{a}\circ(\ee_{b}\circ(\ee_{c}\circ\ee_{d}))=\pm 1\}$,  
makes the equality $A\cv_X B= A\cv B$ to hold {\color{black}{for such particular values of $X$}} \cite{dix}.

Eq.~(\ref{3p14}) shows that {\color{black}{the octonionic field}} $X\in \sec(\mathbb{O}S^7)$ determines two endomorphisms of the octonionic algebra, $f_1, f_2\in$ End($\sec(\mathbb{O}S^7)$), defined by $
A \circ_X B = f_1(A \circ f_1^{-1}(B)) = f_2(f_2^{-1}(A) \circ B)$, for all $A, B \in \sec(\mathbb{O}S^7).$ The \textit{quasi-alternativity} of the $\circ_X$-multiplication then follows as
\begin{gather}
A\circ_X (A\circ_X B)=(A\circ A)\circ_X B,  \qquad (A\circ_X B)\circ_X B=A\circ_X (B\circ B).
\end{gather}
The $X$-product can be, thus, seen as the original octonionic product. In fact, 
 there exists an orthogonal mapping $T\in {\rm SO}(\mathbb{R}^{0,7})$, such that the mapping $\rho: (V_7, \circ)=\mathbb{O}\to(V_7, \circ_X) = \mathbb{O}_X$, given by $a+ \vv \overset{\rho}{\mapsto} a+ T(\vv)$, is an isomorphism, for all $a\in \mathbb{R}$ and $\vv \in \mathbb{R}^{0,7}$ \cite{daRocha:2012tw}. The reciprocal statement is up to now a conjecture.
Besides, an orbit whose elements are isomorphic copies of $\mathbb{O}$ obtained out of any fixed copy of $\mathbb{O}$ is an orbifold  
$ S^7/\ZZ_2=\mathbb{R} \mathrm{P}^7$, being diffeomorphic to SO(7)/$G_2$. In fact, identifying two antipode points on $S^7$ yields  $A\cv_{-X}B=A\cv_X B$. One of the most natural ways of obtaining a parallelizable $S^7$ is choosing two non-canonical connections on Spin(7)/$G_2$ \cite{Lukierski:1983qg}. 

Besides, the sphere $S^7$ plays a prominent role on the (quaternionic\footnote{We denote hereon by $\mathbb{H}$ the ring of quaternions.}) Hopf fibration $
S^3\hookrightarrow S^7\overset{p}{\rightarrow} S^4$,  \cite{jayme}.
In this sense, $S^7$ can be realized as being the set $\{(q_1,q_2)\in \mathbb{H}^2\,|\,\|q_1\|^2+\|q_2\|^2=1\}$, where $p:S^7\to S^4$ maps the pair $(q_1,q_2)$ to $q_1/q_2$, an element in the projective line $\mathbb{H P}^1\approx S^4$. Thus, each fiber is represented by a torsor
 that is parametrized by quaternions of unit norm, defining $S^3$. 
A construction of this Hopf algebra was also realized using regular spinors, being the most important realization with respect to the Lounesto's spinor field classification, in Refs. \cite{jayme,daRocha:2008we,EPJC}. 

More generally speaking, without considering just the $S^7$ manifold, a $n$-manifold $\mani$ is said to have the property of global parallelizability if there are $n$ linearly independent vector fields defined on $\mani$. Thereupon, for each $X\in\mani$, one can linearly combine these fields to obtain an orthonormal basis for $T_XM$. Given one of these bases, since vectors are linear combinations of such elements, their covariant derivative in different points can be taken in a natural way, which results in path independence for the parallel transport. In fact, it follows that
\bege
[\mathring{D}_{\mu},\mathring{D}_{\nu}]=0=R_{\mu\nu},
\enge
where $\mathring{D}$ denotes the covariant derivative defined with respect to this parallel transport,  {\color{black}{whereas $R_{\mu\nu}$ denotes the curvature tensor}}. As usual, $
\mathring{D}=\partial+\mathring{\Gamma}=D-T,$ 
where $T$ denotes the parallelizing torsion and $\mathring{\Gamma}$ is the parallelizing connection. Let ${e_{\nu}}^{\rm a}$ indicate the \textit{vielbein}, related to a non-coordinate basis, wherein roman letters indicate the indexes of the tangent spaces, accordingly. As $D_{\mu}{e_{\nu}}^{\rm a}=0$, the covariant derivative of the \textit{vielbein} yields 
$
\mathring{D}_{\mu}{e_{\nu}}^{\rm a}={-T_{\mu\nu}}^{\rm a}
$. Now, one can look at a manifold $\mani$, that for our case is $S^7$, and consider the infinitesimal translations determined by the covariant derivatives. As may be seen, it is straightforward that these translations configure a closed algebra \cite{ced}:
\bege
[D_{\rm a},D_{\rm b}]=[{e_{\rm a}}^{\mu}D_{\mu},{e_{\rm b}}^{\nu}D_{\nu}]=2{e_{\rm a}}^{\mu}[D_{\mu},{e_{\rm b}}^{\nu}]D_{\nu}=2{e_{\rm a}}^{\mu}{T_{\mu \rm b}}^{\nu}D_{\nu}=2{T_{\rm ab}}^{\rm c}D_{\rm c}.
\enge
 When the manifold is also a group manifold, it is evident that the parallelizing torsion does not depend on the point chosen and is, thus, only expressed by the structure constants. Nonetheless, $S^7$ must be carefully considered, for the torsion varies at each point on  the manifold. This fact is intrinsically related to the non-associativity of $\oct$, as it can be seen in Ref. \cite{ced}.

For {\color{black}{a field}} $X\in\sec(\oct S^7)$, one can construct a parametrization of $S^7$ with respect to unitary octonionic fields {\color{black}{$\frac{X}{|X|}\in\sec(\oct S^7)$}}. The tangent space $T_XS^7$ is spanned by the usual octonionic basis as $\{X\circ \ee_i\}_{i = 1}^7$. Now, as introduced in {\color{black}{Ref. }}\cite{ced}, let us consider  the infinitesimal operator $\delta_A$, where $A\in\sec(\oct S^7)$ is now a pure imaginary octonionic field, acting on $X$ as $\delta_A X=X\circ A$. This transformation defines the parallel transport on the basis spanned by the choice of $X$. An explicit derivation can be realized \cite{ced} to find the commutator of the defined transformations:
\begin{eqnarray}
[\da,\db]\x &\equiv& \da(\db \x)-\db(\da \x)
	    =\x\circ\bigl(\bar{X}\circ((X\circ\,B\,)\circ A)-\bar{X}\circ((X\circ A)\circ\,B\,)\bigr).
\end{eqnarray}
It can be shown that the parameter
$
\bar{X}\circ((X\circ\,B\,)\circ A)-\bar{X}\circ((X\circ A)\circ\,B\,)\is 2\{\bar{X}\circ((X\circ\,B\,)\circ A)\}
$  
is twice the negative of the parallelizing torsion \cite{roo}. Componentwise, 
\bege\label{123123}
T_{\rm abc}(X)=[(\bar{\ee}_{\rm a}\circ\bar{X})\circ(X\circ \ee_{\rm b})\circ \ee_{\rm c}] \quad  \mbox{and} \quad  [\da,\db]=2T_{\rm abc}(X)\delta_{\rm c},
\enge
presenting, thus, a Moufang loop (or Moufang
quasigroup) structure in the second equation in (\ref{123123}). Therefore, one can see that the operator $\delta$ and the parallelizing covariant derivative are, in fact, in a 1-1 correspondence. Now, taking another field $\zeta\in\oct$, with ${\zeta\over |\zeta|}\is Y\in \sec(\oct S^7)$, over the same orientation given by the choice of $X$, and transforming it such that  the relations in Eq. (\ref{123123}) are preserved, such properties preclude the straightforward \emph{ansatz} $\da Y\is Y \circ \,A\,$ \cite{ced}. The two regarded fields on $S^7$ must, thus, transform according to another rule, that may seem at a first glance, not the simplest choice. Ref. \cite{ced} derived the appropriate 
transformation rule for fermionic fields on $S^7$, taking into account its  underlying parallelizable torsion, as $
\delta_AY=Y\circ_X A.$

Now, the new classes of spinors on $S^7$ can be lifted onto the parallelizable $S^7$. 
In fact, for it we need to remember the equivalence between the classical and the algebraic spinor fields. Going back to the 4D Minkowski spacetime, the standard Dirac spinor $\psi$ was identified, e. g., in Ref. \cite{lou2} as an element of the minimal left ideal $(\mathbb{C}\otimes \cl _{1,3}){\rm f}$ associated to the Dirac--Clifford algebra $(\mathbb{C}\otimes \cl _{1,3})$, generated by the primitive idempotent 
$
{\rm f}=\frac{1}{4}(1+\gamma_{0})(1+i\gamma_{1}\gamma_{2})$ yielding $\psi \in (\mathbb{C}\otimes \cl _{1,3})f$ is an algebraic spinor \cite{lou2}. Hence, using the Dirac representation of the gamma matrices, the algebraic spinor\begin{equation}
\psi =
\begin{pmatrix}
\psi_1 & 0 & 0 & 0 \\
\psi_2 & 0 & 0 & 0 \\
\psi_3 & 0 & 0 & 0 \\
\psi_4 & 0 & 0 & 0
\end{pmatrix}\in (\mathbb{C}\otimes \cl _{1,3}){\rm f} \simeq \mathcal{M}(4,\mathbb{C}){\rm f}, \end{equation}  is equivalent to the classical spinor $\psi=(\psi_1, \psi_2, \psi_3, \psi_4)^\intercal
\in \mathbb{C}^{4}.$ 

Now, this concept can be extended for the parallelizable $S^7$, emulating the transformation $\delta_A \psi =\psi\circ_X A$ that can  encompass algebraic spinor fields.
For it, let us consider the  
Clifford algebra $\cl_{0,7}$ on a tangent space $T_X S^7$, at a point $X\in S^7$. According to the Radon--Hurwitz theorem in the Appendix \ref{app1},  for 
$k = 7 - r_7 = 4$, one aims the set $\{e_{I_1}, \ee_{I_2}, \ee_{I_3}, \ee_{I_4}\}\subset \cl_{0,7}$ that commute and  squares the identity \cite{Vaz:2016qyw}. Identifying, for  example  \cite{daRocha:2012tw}  $\ee_{I_1} = \ee_{1}\ee_{2}\ee_{3}$, 
$\ee_{I_2} = \ee_{1}\ee_{4}\ee_{5}$, $\ee_{I_3} = \ee_{1}\ee_{6}\ee_{7}$, and  
$\ee_{I_4} = \ee_{3}\ee_{4}\ee_{7}$, yields that the idempotent 
\begin{equation*}
f = \frac{1}{16}(1+\ee_1\ee_2\ee_3) 
(1+\ee_1\ee_4\ee_5)
(1+\ee_1\ee_6\ee_7) 
(1+\ee_3\ee_4\ee_7)\in\cl_{0,7}
\end{equation*}
is a primitive one. 
Hence, a spinor $\psi\in S^7$ has its algebraic version as the  element
$\mathring\psi f$ of the left ideal $\cl_{0,7}f$, for some multivector $\mathring\psi\in\cl_{0,7}$. This is accomplished just for introducing the 
$S^7$ spinor into the Clifford bundle itself, on $S^7$.

Now, to write the correct transformation of a fermionic field 
on the parallelizable $S^7$, given an element of the vector space underlying  $\cl_{0,7}$, a non-associative product called the $\xi$-product was introduced in \cite{JA} as a natural generalization for the $X$-product. For homogeneous multivectors $\xi=u_1\wedge\ldots \wedge u_k\in\sec\la^k(\mathbb{R}^{0, 7})\hookrightarrow\sec\cl_{0, 7}$, where $\left\{u_p\right\}^{k}_{p=1}\subset\sec T\mathbb{R}^{0, 7}$  and $A\in\sec(\mathbb{O}S^7)$, the products $\bullet_\llcorner$ and $\bullet_\lrcorner$ are  defined (and extended by linearity) by \cite{JA,daRocha:2012tw}
\beq \bullet_\llcorner \colon \sec(\mathbb{O}S^7)\times \sec\la^k(\mathbb{R}^{0, 7})&\to& \sec(\mathbb{O}S^7)\nonumber\\
(A, \xi)&\mapsto& A\underset{\tiny{X}}{\bullet_\llcorner} \xi=(\cdots((A\circ_X u_1)\circ u_2)\circ \cdots)\circ u_k, \label{209}\\
\bullet_\lrcorner \colon \sec\la^k(\mathbb{R}^{0, 7})\times \sec(\mathbb{O}S^7)&\to&\sec(\mathbb{O}S^7)\nonumber\\
(\xi, A)&\mapsto& \xi\underset{\tiny{X}}{\bullet_\lrcorner} A=u_1\circ (u_2\cv(\cdots \circ (u_k \circ A))\cdots ).\label{210}\eeq

Hence, within the above constructions, the transformation of the reconstructed spinor field on $S^7$  from its bilinear covariants in Eq. (\ref{3}), that is a representative of the new classes (\ref{c12} -- \ref{c14}) of spinor fields on $S^7$, can be defined as
\beq\label{ferm1}
\delta_A\psi=\psi\underset{\tiny{X}}{\bullet_\lrcorner} A, \quad \forall A\in\sec(\mathbb{O}S^7).
\eeq
In this way, the previous new classes of $S^7$ spinors are lifted
onto the parallelizable $S^7$. This transformation is compatible to the ones defined in Ref. \cite{top}.

\section{Conclusions}

We have managed to establish the reconstruction theorem for the new classes of spinor fields on $S^7$ using the generalized Fierz aggregate, for each recently found new class of spinor fields on the $S^7$ spin bundle according to their bilinear covariants. Besides, this categorization has enabled the construction of new fermionic fields on the parallelizable  $S^7$, promoting the new classes of classical spinor fields on $S^7$ to new classes of algebraic ones. Hence, the correct transformation of these elements, generating a Moufang loop structure on the parallelizable $S^7$ was derived.  Aiming to this procedure, we briefly reviewed the parallelizability property on the parallelizable $S^7$, wherein the parallel transport could be analyzed with respect to the torsion. Therein, the non-associativity of the octonionic bundle on $S^7$ was related to  the torsion tensor on the parallelizable $S^7$, as a function dependent on each point on $S^7$, via the $X$-product. In this way, additional classes of fermionic (spinor) fields on the parallelizable $S^7$ have been constructed, according to the classes obtained heretofore, lifted from the $S^7$ spin bundle, with the right transformation under infinitesimal transformations. Our results, thus, generalize the ones in Ref. \cite{ced}, also proposing new classes of fermionic fields that may play the role of the solutions in compactfications of supergravity. 

\subsubsection*{Acknowledgments}
AYM thanks to FAPESP (grant No. 2016/14021-8) and RdR~is grateful to CNPq (Grant No. 303293/2015-2),
and to FAPESP (Grant No.~2017/18897-8), for partial financial support.

\appendix
{\color{black}{\section{The spinor bundle}
\label{app}
Here the spinor bundle of Minkowski spacetime is introduced. Such a structure 
can be emulated for any spacetime $(p,q)$ signature, when the regarded manifold $M$ has a spin structure. In particular, the construction for the $(0,7)$ and $(7,0)$ signatures, regarding the $S^7$ spinor bundle, are similarly constructed. As the Minkowski spacetime is the most illustrative and phenomenologically explored, we want to fix the notation
and the intuitive setup for the spinor bundles. 
     
     In what follows, one denotes the connected component to the identity of the Spin group by $\mathrm{Spin}_{1,3}^{\rm e}\simeq\mathrm{SL}(2,\mathbb{C)}$, being the universal covering group of the (restricted) Lorentz group $\mathrm{SO}_{1,3}^{\rm e}$. Again we denote by $M$ the Minkowski spacetime $\mathbb{R}^{1,3}$. 
Given the Minkowski metric tensor and given the 
 principal bundle of frames on the manifold $M$,  the orthonormal frame [coframe] 
bundle shall be denoted by $\mathbf{P}%
_{\mathrm{SO}_{1,3}^{\rm e}}(M\mathbf{)}$ [$P_{\mathrm{SO}_{1,3}^{\rm e}}(M)]$. Considering $M$ a spin manifold, there  exists the spin frame and the spin coframe bundles, respectively denoted by  $\mathbf{P}_{\mathrm{Spin}%
_{1,3}^{\rm e}}(M\mathbf{)}$ and
$P_{\mathrm{Spin}_{1,3}^{\rm e}}(M\mathbf{)}$. 
Sections of $P_{\mathrm{SO}_{1,3}^{\rm e}%
}(M\mathbf{)}$ and of $P_{\mathrm{Spin}%
_{1,3}^{\rm e}}(M\mathbf{)}$ are both orthonormal coframes.
However, contrary to the $P_{\mathrm{SO}_{1,3}^{\rm e}%
}(M\mathbf{)}$ bundle,  coframes  on the bundle $P_{\mathrm{Spin}_{1,3}^{\rm e}}(M\mathbf{)}$ that differ by a $2\pi$ rotation are considered to be distinct, whereas 
coframes that can be led into each other by a $4\pi$ rotation are equivalent.

The 
fundamental mapping  
$s:P_{\mathrm{Spin}_{1,3}^{\rm e}}(M\mathbf{)\rightarrow}P_{\mathrm{SO}_{1,3}^{\rm e}%
}(M\mathbf{)}$ completely defines the bundle   $P_{\mathrm{Spin}_{1,3}^{\rm e}}(M\mathbf{)}$. A spin structure on
$M$ consists of a principal fiber bundle, endowed with a canonical projection 
$\mathbf{\pi}_{s}:P_{\mathrm{Spin}_{1,3}^{\rm e}}(M)\rightarrow M$, under the conditions:

(i) Given the projection mapping $\pi: P_{\mathrm{SO}%
_{1,3}^{\rm e}}(M)\to M$,   then $\mathbf{\pi}(s(p))=\mathbf{\pi}_{s}(p),$ for all element $p\in P_{\mathrm{Spin}%
_{1,3}^{\rm e}}(M)$.

(ii) Denoting by $\mathrm{Aut}(\cl_{1,3})$ the set of automorphisms (namely, isomorphisms from  $\cl_{1,3}$ to itself), given the adjoint mapping 
\begin{eqnarray}
\mathrm{Ad}:\mathrm{Spin}_{1,3}^{\rm e}&\rightarrow&\mathrm{Aut}(\cl_{1,3})\nonumber\\
\tau&\mapsto&\mathrm{Ad}_{\tau}: \cl_{1,3}\to \cl_{1,3}\nonumber\\
 &&\quad\qquad\quad\xi\mapsto \tau\xi\tau^{-1},
 \end{eqnarray}
 then $s(p \tau)=s(p)\mathrm{Ad}_{\tau},$ for all element $p\in
P_{\mathrm{Spin}_{1,3}^{\rm e}}(M)$.

 The Clifford bundle of differential forms
$\mathcal{C\ell
(}M,g)$ is a vector bundle associated with $P_{\mathrm{Spin}%
_{1,3}^{\rm e}}(M\mathbf{)}$, whose sections are sums of
non-homogeneous differential forms. Hence 
$\mathcal{C\ell(}M,g)\simeq P_{\mathrm{SO}_{1,3}^{\rm e}}(M)\times
_{\mathrm{Ad}^{\prime}}\cl_{1,3}$ is a bundle defined by:

(1) Let $\mathbf{\uppi}:\mathcal{C}\ell(M,g)\rightarrow M$ be
the canonical projection and let
$\{U_{\alpha}\}$ be an open covering of $M$. There are
trivialization mappings
$\mathbf{\psi}_{i}:\mathbf\uppi^{-1}(U_{i})\rightarrow U_{i}%
\times\cl_{1,3}$ of the form $\mathbf{\psi}_{i}(p)=(\uppi(p),\psi_{i,x}(p))=(x,\psi_{i,x}(p))$. If $x\in U_{i}\cap
U_{j}$ and $p\in\mathbf\uppi^{-1}(x)$, then $
\psi_{i,x}(p)=h_{ij}(x)\psi_{j,x}(p)$, 
for $h_{ij}(x)\in\mathrm{Aut}(\cl_{1,3})$, where $h_{ij}:U_{i}\cap
U_{j}\rightarrow\mathrm{Aut}(\cl_{1,3})$ are the transition
mappings of $\mathcal{C}\ell(M,g)$. It is worth to emphasize that every
automorphism of $\cl_{1,3}$ can be written as $
h_{ij}(x)\psi_{j,x}(p)=a_{ij}(x)\psi_{i,x}(p)a_{ij}(x)^{-1}$ for some invertible
element of $a_{ij}(x)\in\cl_{1,3}$.
In other words, 
$\mathrm{Ad}_{a_{ij}}=h_{ij}$ in all intersections $U_{i}\cap
U_{j}$.

(2) Besides, when the adjoint mapping is restricted to the group ${\mathrm{Spin}_{1,3}^{\rm e}}$, it 
defines the mapping $\mathrm{Ad}|_{\mathrm{Spin}_{1,3}^{\rm e}}: \mathrm{Spin}_{1,3}^{\rm e}\rightarrow\mathrm{SO}_{1,3}^{\rm e}$, with kernel $\mathbb{Z}_{2}$. Hence $\mathrm{Ad}:\mathrm{Spin}_{1,3}^{\rm e}\rightarrow\mathrm{Aut}%
(\cl_{1,3})$ descends to a representation $\mathrm{Ad}%
^{\prime}:\mathrm{SO}_{1,3}^{\rm e}\rightarrow\mathrm{Aut}(\cl_{1,3})$ of
$\mathrm{SO}_{1,3}^{\rm e}$, yielding $\mathrm{Ad}_{\sigma(\tau)}^{\prime}\xi=\mathrm{Ad}_{\tau}%
\xi=\tau\xi\tau^{-1}$.

(3) The main group underlying the Clifford bundle
$\mathcal{C}\ell(M,g)$ is, thus, reducible to $\mathrm{SO}_{1,3}^{\rm e}$. The transition mappings of  $P_{\mathrm{SO}_{1,3}^{\rm e}%
}(M)$ can be then regarded  as
transition mappings of the Clifford bundle, yielding  \cite{moro,lawmi}
\begin{equation}
\mathcal{C}\ell(M,g)=P_{\mathrm{SO}_{1,3}^{\rm e}}(M)\times
_{\mathrm{Ad}^{\prime}}\cl_{1,3}=P_{\mathrm{Spin}_{1,3}^{\rm e}}(M)\times
_{\mathrm{Ad}}\cl_{1,3}. 
\end{equation}
Hence, spinor fields are sections of vector bundles associated with  the
principal bundle of spinor coframes. The well known regular Minkowski spinor
fields are sections of the bundle
\begin{equation}
S=P_{\mathrm{Spin}_{1,3}^{\rm e}}(M)\times_{\rho}\mathbb{C}^{4}, \label{4.7}%
\end{equation}
with $\rho$ being the $D^{(1/2,0)}\oplus D^{(0,1/2)}$ representation of $\mathrm{Spin}%
_{1,3}^{\rm e}$ onto the space of linear mappings on $\mathbb{C}^{4}$.

With what was exposed heretofore in the Appendix A, one straightforwardly introduces, \emph{mutatis mutandis}, the analogous underlying spinor bundle on the 7-sphere.}}

\section{The Radon--Hurwitz theorem}
\label{app1}
Let $\cl_{p,q}$ 
be the Clifford algebra associated to $\mathbb{R}^{p,q}$ and
$\{\ee_i\}$ $(i=1,\ldots,n)$ an orthonormal basis of this  
quadratic space. A primitive idempotent of $\cl_{p,q}$ is given by $f = \frac{1}{2}(1+\ee_{I_1})\cdots \frac{1}{2} 
(1+\ee_{I_k})$, 
where $\{\ee_{I_1},\ldots,\ee_{I_k}\}$ is a set of
elements in $\cl_{p,q}$ that commute and such that 
$(\ee_{I_\alpha})^2 = 1$ for $\alpha = 1,\ldots,k$. It generates a group of order  $2^k$, where $k = q - r_{q-p}$, and 
$r_j$ are the Radon--Hurwitz numbers defined by \cite{Vaz:2016qyw} \index{Radon--Hurwitz!numbers}
\begin{table}[h]
{}
{
\begin{tabular}{||c||c|c|c|c|c|c|c|c||} \hline\hline
$j$ & 0 & 1 & 2 
& 3 & 4 & 5 & 6 & 7 \\ \hline
$r_j$ & 0 & 1& 2& 2& 3 & 3 & 3& 3 \\\hline\hline
\end{tabular}
}
\end{table}
with the  recurrence relation $r_{j+8} = r_j + 4$.

\end{document}